\documentclass[twocolumn]{aastex62}
\usepackage[latin9]{inputenc}

\shorttitle{Probing Cosmic Ray Transport}
\shortauthors{Thomas, Pfrommer \& En{\ss}lin}

\begin{document}

\title{Probing Cosmic Ray Transport with {\em Radio Synchrotron Harps} in the Galactic Center}

\correspondingauthor{Timon Thomas}
\email{tthomas@aip.de}

\author{Timon Thomas}
\affil{Leibniz-Institute for Astrophysics Potsdam (AIP), An der Sternwarte 16, 14482 Potsdam, Germany}

\author{Christoph Pfrommer}
\affiliation{Leibniz-Institute for Astrophysics Potsdam (AIP), An der Sternwarte 16, 14482 Potsdam, Germany}

\author{Torsten En{\ss}lin}
\affiliation{Max Planck Institute for Astrophysics, Karl-Schwarzschild-Str. 1, 85741 Garching, Germany}
 
\begin{abstract}
Recent observations with the MeerKAT radio telescope reveal a unique population of faint non-thermal filaments pervading the central molecular zone (CMZ). Some of those filaments are organized into groups of almost parallel filaments, seemingly sorted by their length, so that their morphology resembles a harp with radio emitting ``strings''. We argue that the synchrotron emitting GeV electrons of these radio harps have been consecutively injected by the same source (a massive star or pulsar) into spatially intermittent magnetic fiber bundles within a magnetic flux tube or via time-dependent injection events. After escaping from this source, the propagation of cosmic ray (CR) electrons inside a flux tube is governed by the theory of CR transport. We propose to use observations of radio harp filaments to gain insight into the specifics of CR propagation along magnetic fields of which there are two principle modes: CRs could either stream with self-excited magneto-hydrodynamical waves or diffuse along the magnetic field. To disentangle these possibilities, we conduct hydrodynamical simulations of either purely diffusing or streaming CR electrons and compare the resulting brightness distributions to the observed synchrotron profiles of the radio harps. We find compelling evidence that CR streaming is the dominant propagation mode for GeV CRs in one of the radio harps. Observations at higher angular resolution should detect more radio harps and may help to disentangle projection effects of the possibly three-dimensional flux-tube structure of the other radio harps.
\end{abstract}


\section{Introduction} \label{sec:introduction}

Radio observations of the Galactic center region show many isolated, elongated filaments \citep{1999Lang,2001LaRosa,2004Nord,2004YusefZadeh}. Recent high-resolution observations with the MeerKAT radio telescope found that the filaments trace bipolar bubbles that are rising from the CMZ near the Galactic center \citep{2019Heywood}. The filaments are characterised by a high aspect ratio, a filament-aligned magnetic field \citep{1999Lang}, strongly polarized emission \citep{2001LaRosa}, and a hard spectral index that steepens away from the geometric center of the filaments \citep{2008Law}. Hence, these non-thermal filaments (NTFs) are illuminated by synchrotron-emitting electrons.

Explanations for injecting relativistic electrons into NTFs include magnetic reconnection \citep{1992Lesch,2001Bicknell}, acceleration in young stellar clusters \citep{2003YusefZadeh}, magnetized wakes of molecular clouds \citep{1999Shore,2002Dahlburg}, pulsar wind nebula \citep{2017Bykov,2019Barkov}, stellar winds of massive stars \citep{1996Rosner,2019YusefZadeh}, and even annihilation of light dark matter \citep{2011Linden}. Whether the origin of the parsec-sized straight NTFs is causally linked to the electron source that powers them is unclear. 

\begin{figure}
    \includegraphics[width=\columnwidth]{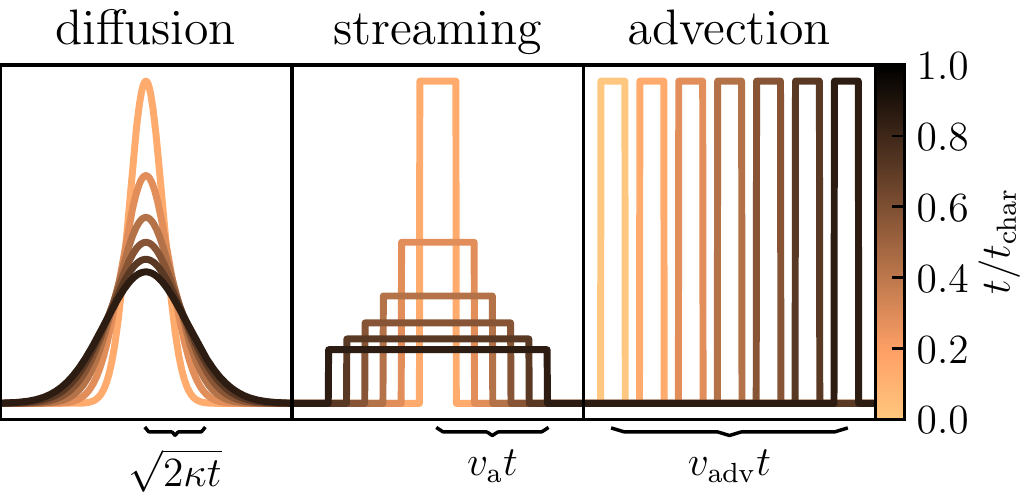}
    \caption{Archetypical transport modes of CR fluids. Left: when CRs are weakly scattered by Alfv\'en waves, they diffuse away from a given source (after an initial time) with a root-mean-square velocity of $\sqrt{2\kappa/t}$ (where $t$ is the time and $\kappa$ denotes the diffusion coefficient) along the magnetic field. Middle: if CRs are effectively scattered, they stream with the Alfv\'en speed, $v_a$, along the magnetic field. Either one or both of these processes may be realized while CRs are tied to frozen-in magnetic fields, causing them to be advected with the bulk plasma velocity, $v_\mathrm{adv}$ (right). }
    \label{fig:kernel_profiles}
\end{figure}

To explain the brightness of NTFs, we need to take a closer look at CR propagation. The Lorentz force ties CRs to any macroscopic magnetic field and causes the CRs to follow the field line motion. When magnetic fields are frozen into and move along with the fluid, CRs are bound to follow these fluid motions. This is called \textit{CR advection} and shown in the right-hand panel of Fig.~\ref{fig:kernel_profiles}. We expect CR advection to be unimportant for NTFs as their straight morphology excludes large-scale gas motions perpendicular to the NTFs that change their appearance.

\begin{figure*}
    \centering
    \includegraphics[width=0.77\textwidth]{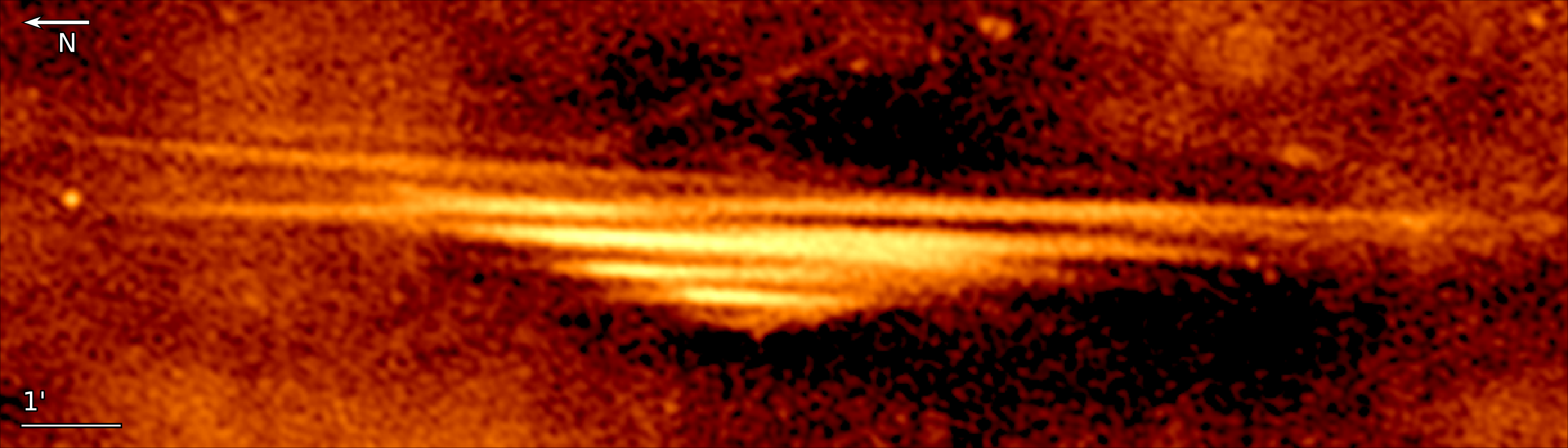}
    \includegraphics[width=0.22\textwidth]{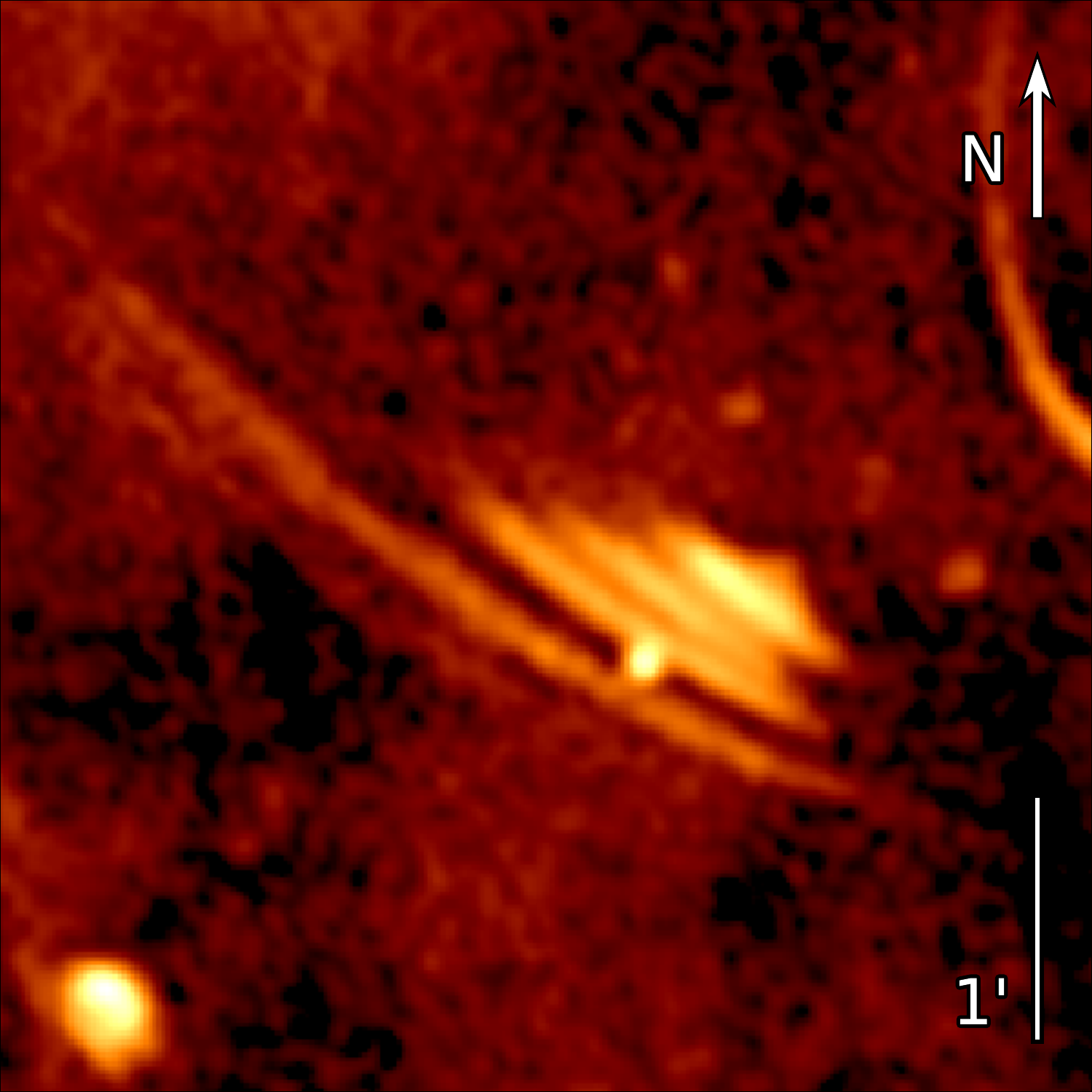}
    \caption{Two radio harps in the MeerKat observation of the CMZ \citep{2019Heywood}. Left: the NTF G359.85+0.39 was discovered by \citet{2001LaRosa}; also named N10 in \citet{2008Law}. Right: G359.47+0.12 was first imaged by \citet{2019Heywood}. Their names correspond to their position in Galactic coordinates.}
    \label{fig:harps}
\end{figure*}

Of particular interest for NTFs is CR propagation along the mean magnetic field. It can be classified into two principle modes depending on the frequency of particle scatterings with magneto-hydrodynamic waves. (i) If these scatterings are frequent, then the ensemble average of the particle distribution follows the motions of their non-relativistic scattering centers while individual particles move with their relativistic velocities. This is the basis for describing CRs as a hydrodynamical fluid on scales larger than the effective mean-free path. (ii) If CR-wave scatterings are infrequent, CRs move ballistically and a kinetic description of individual particle trajectories is appropriate. \citet{2017Malkov} showed that CRs leave the ballistic regime after three characteristic scattering times and enter a diffusive, fluid-like behaviour (left panel of Fig.~\ref{fig:kernel_profiles}).

For CRs with energies below $\sim200$ GeV, magneto-hydrodynamical (MHD) waves are believed to be the dominant source of scattering \citep{2011Yan,2012Blasi}. CRs can provide their own scattering centers by exiting Alfv\'en waves on scales comparable to their gyroradii through the gyroresonant instability \citep{1969Kulsrud}. These Alfv\'en waves interact with CRs so that the effective CR drift velocity approaches the Alfv\'en velocity, $v_a$, which is referred to as \textit{CR streaming}. 

CRs injected by a compact source excite Alfv\'en waves while leaving their acceleration site. These Alfv\'en waves are travelling in opposite directions along the magnetic field away from the source. Both leading fronts of Alfv\'en waves span an expanding region populated by CRs. Due to their confinement into this region, the CR population rarefies. Assuming perfect confinement, there is a sharp transition between locations that are occupied by or free of injected CRs (see middle panel of Fig.~\ref{fig:kernel_profiles}).

This fundamental difference between CR streaming and diffusion allows us to differentiate between the two modes by studying the radio synchrotron brightness along NTFs: (i) the synchrotron emission from diffusing CR electrons smoothly fades away from a compact source while streaming CR electron populations show a central constant brightness level and a sharp transition to any background emission  and (ii) as indicated in Fig.~\ref{fig:kernel_profiles}, the root-mean-square distance of diffusing CR electrons increases as $\sqrt{2\kappa t}$ while in the CR streaming model, it increases linearly with time as $v_\mathrm{a}t$. If we were to observe equidistantly-spaced snapshots of the two propagation modes, then the envelope of the snapshots should either show a bell shape (for CR diffusion), a triangle (for pure CR streaming), or an inverse bell shape (for CR streaming~+~diffusion).

In this Letter we are studying a particular class of NTFs that we call {\em radio synchrotron harps} and of which we show two examples in Fig.~\ref{fig:harps}. We will argue that those objects provide a rich avenue to study CR transport and propagation using radio observations.

\section{Sources Powering Non-Thermal Filaments}
\label{sec:sources}

A massive star or pulsar moving through the CMZ with velocity $v_*\sim{v}_\mathrm{a}$ can intersect and inject CRs into a magnetic flux tube that has been stretched by the bipolar outflow from the CMZ \citep{2019Heywood}. We conjecture that the perpendicular radio harp sizes correspond to the radial flux-tube extents while the regular arrangement of the harp ``strings'' in Fig.~\ref{fig:harps} either represents intermittent magnetic fiber bundles within a magnetic flux tube or are a signature of time-dependent injection of CRs into the more homogeneous magnetic flux tube. In the latter case, the intermittency of the radio ``strings'' might reveal details of the magnetic reconnection process around the wind termination shock that enables CRs to escape. In both cases, the different ``string'' lengths show a chronological sequence of CR injection events onto an NTF. After injection, the CRs propagate along the magnetic filament, which decreases their energy density and increases their spatial extent. Hence, NTFs with more recently injected CRs appear shorter and brighter while previously injected CRs form longer and fainter filaments. The resulting morphology is that of a filamentary isosceles triangle (or bell) with a bright apex and a fainter base, see Fig.~\ref{fig:harps}.

\paragraph{Wind termination shocks of massive stars} Electrons generated at wind termination shocks or bow shocks of massive stars can illuminate NTFs \citep{1996Rosner,2019YusefZadeh}. Massive stellar winds interact with their local interstellar medium (ISM) by building up an interaction layer between the wind interior and the ISM. This layer is confined by a bow shock that encompasses the shocked ISM and a wind termination shock. These shocked fluids are initially separated by a contact discontinuity, which becomes unstable due to Rayleigh-Taylor instabilities that cause mixing of both fluids. At both shocks, low-energy electrons can be accelerated to relativistic energies via diffusive shock acceleration \citep[e.g.,][]{2018delValle}. Some bow shock complexes are luminous enough for observable synchrotron emission \citep[][for the bow shock of a runaway O star]{2010Benaglia}. The stand-off radius $R$ between star and bow shock is given by the pressure balance between stellar wind and ISM: 
\begin{equation}
    R=\left(\frac{\dot{M}v_\infty}{4\pi(\rho_{\mathrm{ISM}}v^2_\star+P_\mathrm{ISM}+B^2/8\pi)}\right)^{1/2}\sim0.05~\mathrm{pc},
\end{equation}
where $\dot{M}\sim(10^{-8}$--$10^{-5})\mathrm{M}_{\odot}~\mathrm{yr}^{-1}$ is the mass loss rate, $v_\infty\sim(1000$--$2500)~\mathrm{km}~\mathrm{s}^{-1}$ is the terminal wind velocity, $v_\star\sim\mathrm{few}\times10~\mathrm{km}~\mathrm{s}^{-1}$ is the relative velocity of the star, $\rho_{\mathrm{ISM}}$ and $P_\mathrm{ISM}$ are the ambient ISM density and pressure, and $B$ is the ISM magnetic field strength. We assume that the NTFs are embedded in the warm CMZ phase with gas temperature $T=10^4~\mathrm{K}$ and number density $n=100~\mathrm{cm}^{-3}$. This implies magnetically dominating NTFs with $B\sim200~\mu$G and a plasma beta $\beta=P_\mathrm{ISM}/(B^2/8\pi)=2c_s/v_\mathrm{a}\sim0.1$, which explains the straight NTF morphology that is not affected by turbulent gas motions. The total kinetic luminosity of the stellar wind is
\begin{equation}
    L_\mathrm{wind}=\frac{1}{2}\dot{M}v^2_\infty\sim1\times10^{35}~\mathrm{erg}~\mathrm{s}^{-1}
\end{equation}
so that the wind termination shock is 
\begin{equation}
\frac{L_\mathrm{wind}}{L_\mathrm{bow}}=\frac{\dot{M}v^2_\infty}{ \rho_\mathrm{ISM}v_\star^3{2}\pi{R}^2}\sim10^2
\end{equation}
times more powerful in comparison to the bow shock, implying that the termination shock dominates the yield of accelerated CRs. Assuming that all kinetic wind energy is dissipated at the wind-termination shock and assuming an electron acceleration efficiency of 0.1\%, the total CR electron luminosity is 
\begin{equation}
    L_e=1\times10^{-3}\,L_\mathrm{wind}\sim1\times10^{32}~\mbox{erg~s}^{-1}.
\end{equation}
Magnetized winds of rotating stars result in perpendicular termination shocks that can accelerate electrons \citep{2019Xu} but not protons \citep{2014Caprioli}. 

\begin{figure*}
    \centering
    \includegraphics[width=0.85\columnwidth]{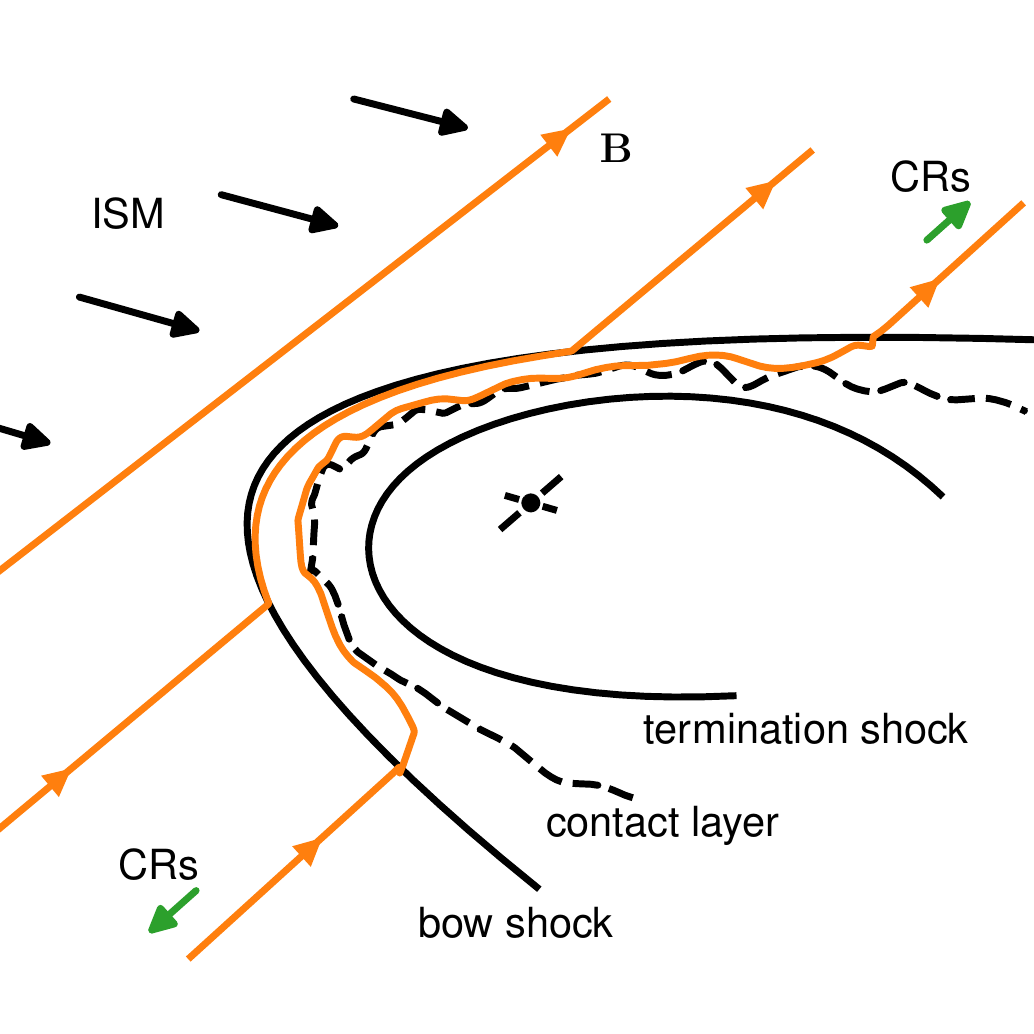}
    \hspace{0.10\columnwidth}
    \includegraphics[width=0.85\columnwidth]{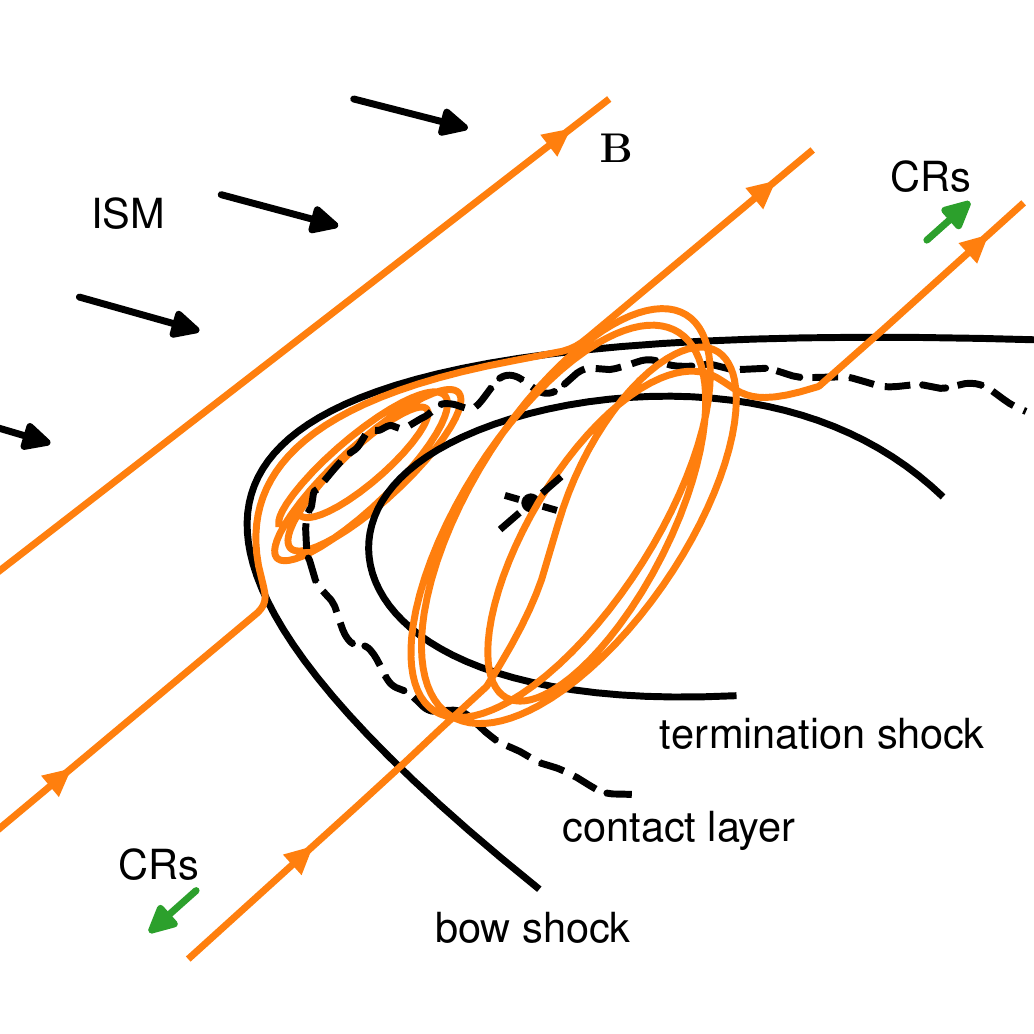}
    \caption{Sketches of possible scenarios that can inject relativistic CRs into NTFs. \textit{Left}: a massive star located in the center drives a stellar wind that builds a wind bubble terminating at a shock. This shock accelerates CRs, which diffuse onto draped ISM magnetic field lines, which experienced mixing with the shocked wind via Rayleigh-Taylor instabilities and escape into the ISM. \textit{Right:} a pulsar drives the wind by accelerating electron-positron pairs towards the wind terminating shock. Piled up field lines behind the wind termination shock can reconnect with the ISM magnetic field, allowing CRs to escape. See text for additional discussion. }
    \label{fig:winds}
\end{figure*}

While moving through the ISM, the stellar wind bubble piles up a magnetic draping layer at the contact discontinuity. Accelerated electrons diffuse onto these field lines and escape from their acceleration site. Subsequently, they move away from the star, emit synchrotron radiation in the strongly magnetized flux tubes of the ISM, and illuminate the NTFs (see left panel of Fig.~\ref{fig:winds}).

\paragraph{Pulsar winds} Another possible source of CR electrons in the NTFs are pulsar wind nebulae \citep[PWN,][]{2019Barkov}. PWNs are fueled by a central pulsar that is characterised by its spin-down luminosity $\dot{E}\sim5\times 10^{37}~\mathrm{erg}~\mathrm{s}^{-1}$. The wind is launched at the light cylinder of the pulsar's magnetosphere, where electron-positron pairs leave the magnetosphere and are accelerated by the strong electromagnetic fields. Further acceleration can take place in reconnection layers of the striped pulsar wind. Similar to a stellar wind, the pulsar wind is separated from the ISM by a layer consisting of the wind termination shock, a contact discontinuity and a possible bow shock. An ISM magnetic field that is draped around the pulsar wind can reconnect at the contact discontinuity with magnetic field originating from the wind interior \citep{2019Barkov,2019BarkovII}. This allows relativistic particles to escape from the PWN into the ISM, see the right-hand panel of Fig.~\ref{fig:winds}. The stand-off distance of the pulsar wind is
\begin{equation}
    R=\left(\frac{\dot{E}}{4\pi{c}(\rho_{\mathrm{ISM}}v^2_\star+P_\mathrm{ISM}+B^2/8\pi)}\right)^{1/2}\sim0.05~\mathrm{pc}.
\end{equation}
Not all electrons leave the PWN so that the luminosity of NTF-injected electron-positron pairs is estimated to be
\begin{equation}
    L_e=2.5\times10^{-4}\frac{\dot{E}}{\sigma}\sim1\times10^{32}~\mathrm{erg}~\mathrm{s}^{-1},
\end{equation}
where $\sigma\sim100$ is the pulsar wind magnetization.

\paragraph{Radio emission from non-thermal filaments} Both scenarios are comparable in terms of their energy budget and size of the acceleration site. Thus, the energy injected into a flux tube
\begin{equation}
E_\mathrm{CR}=f_\mathrm{esc}\frac{R}{v_\star}L_e\sim5\times10^{42}~\mathrm{erg},
\end{equation}
is the same for both sources. Here, $f_\mathrm{esc}\sim0.3-1$ is the time fraction during which CRs near the wind termination shock are injected into a flux tube. Furthermore, assuming that the injected electrons/pairs have a Lorentz factor $\gamma\sim10^3$, they emit synchrotron radiation at 
\begin{equation}
    \nu=\frac{3eB\gamma^2}{2\pi{m_e}c}\sim1.5~\mathrm{GHz}
\end{equation}
with a total luminosity of
\begin{equation}
    L_\mathrm{syn}=E_\mathrm{CR}\frac{\sigma_\mathrm{T}B^2\gamma}{6\pi{m_e}c}\sim2\times10^{29}~\mathrm{erg}~\mathrm{s}^{-1},
\end{equation}
which corresponds to a spectral flux of
\begin{equation}
    F_\mathrm{syn}=\frac{L_\mathrm{syn}}{2\pi{d}^2\nu}\sim2~\mathrm{mJy}    
\end{equation}
at a distance of $d=8.2$ kpc. Within the uncertainties, this matches the radio harp flux. The associated synchrotron cooling time of $\sim10^6$~yr is much longer than the CR propagation time of $\sim60$~kyr so that we do not expect synchrotron fading (see Sect.~\ref{sec:hydro model}).

\section{Hydrodynamic Flux Tube Model for Radio Harps}
\label{sec:hydro model}

Already the detection of radio harps is a strong argument in favor of CR propagation with $v_\mathrm{a}$: CRs leaving the source have individual trajectories that are preferentially aligned with the magnetic flux tube. As NTFs lay mostly perpendicular to the Galactic plane, the synchrotron radiation should be beamed away from the Galactic plane and undetectable for us. Thus, to explain the NTF detection some mechanism is needed that effectively scatters CRs such that their beamed radiation is observable with radio telescopes. A likely possibility is pitch-angle scattering by gyroresonant Alfv\'en waves. CRs moving along a flux-tube can excite these Alfv\'en waves via the gyroresonant instability, which leads to CR streaming close to the Alfv\'en speed, $v_\mathrm{a}$ (see Sect.~\ref{sec:introduction}).

We model CR electron propagation inside NTFs with the following numerical setup: we assume self-similar evolution of the individual filaments in a given harp and that the observation samples the evolution of an archetypical NTF at different times. Within a propagation model, this allows us to conduct a single simulation for all filaments. Filaments of different lengths correspond to different simulation times: longer filaments correspond to later times with a broadened CR distribution. 

We assume an Alfv\'en speed of $v_{\mathrm{a}}=40~\mathrm{km}~\mathrm{s}^{-1}$ and use ISM parameters as detailed in Sect.~\ref{sec:sources}. The simulation domain is aligned with the magnetic flux tube, which is assumed to be straight and to have a constant cross section $\pi{R}^2$ during the simulation. The CR electrons are initialised by injecting $E_\mathrm{CR}=5\times10^{42}~\mathrm{erg}$ into a Gaussian with width $0.05~\mathrm{pc}$ to model CR injection at the bow shock of a massive star or pulsar.

\begin{enumerate}
    \item The \textit{diffusion} model assumes that the CRs diffuse along the magnetic flux tubes with a constant coefficient $\kappa=3\times10^{25}~\mathrm{cm}^2~\mathrm{s}^{-1}$, which was chosen to match NTF sizes with a diffusion length scale $l=\sqrt{2\kappa{t}}$ and $t=30~\mathrm{kyr}$.\footnote{In the \textit{diffusion} model, only the combination $\kappa{t}$ is constrained by the diffusion length; for simplicity, we use the time scale of the \textit{streaming~+~diffusion} model.} We include Alfv\'en wave cooling of CRs \citep[see][]{2017Pfrommer}. 
    \item The \textit{streaming~+~diffusion} model uses the more accurate description for CR transport of \citet{2019Thomas}, which evolves the CR energy and momentum density. In addition, the energy contained in gyroresonant Alfv\'en waves is evolved and coupled to CRs using quasi-linear theory of CR transport. We only consider non-linear Landau damping of Alfv\'en waves \citep[see][]{2019Thomas}. The initial CR energy flux is chosen so that CRs stream with $v_\mathrm{a}$.
\end{enumerate}
The \textit{streaming~+~diffusion} model includes details of the microphysical interaction of CRs and Alfv\'en waves that are absent in the pure \textit{diffusion} model which is unable to model CR streaming. In comparison to the \textit{diffusion} model, where the diffusion coefficient $\kappa$ is constant, the diffusion coefficient in the \textit{streaming~+~diffusion} model is calculated based on the local strength of Alfv\'en waves. We solve the equations of \citet{2019Thomas} using a finite volume method (Thomas et al. in prep) in the moving mesh code \texttt{AREPO} \citep{2010Springel} for both models (in the \textit{diffusion} model $\kappa$ is constant). We use a one-dimensional grid with 4096 cells, a grid spacing of $\Delta x=4\times10^{-3}~$pc, and outflowing boundary conditions. A reduced speed of light $\tilde{c}=1000~\mathrm{km}~\mathrm{s}^{-1}$ is used and we confirmed that the presented results are robust for changes of $\tilde{c}$.
\begin{figure}
    \includegraphics[width=\columnwidth]{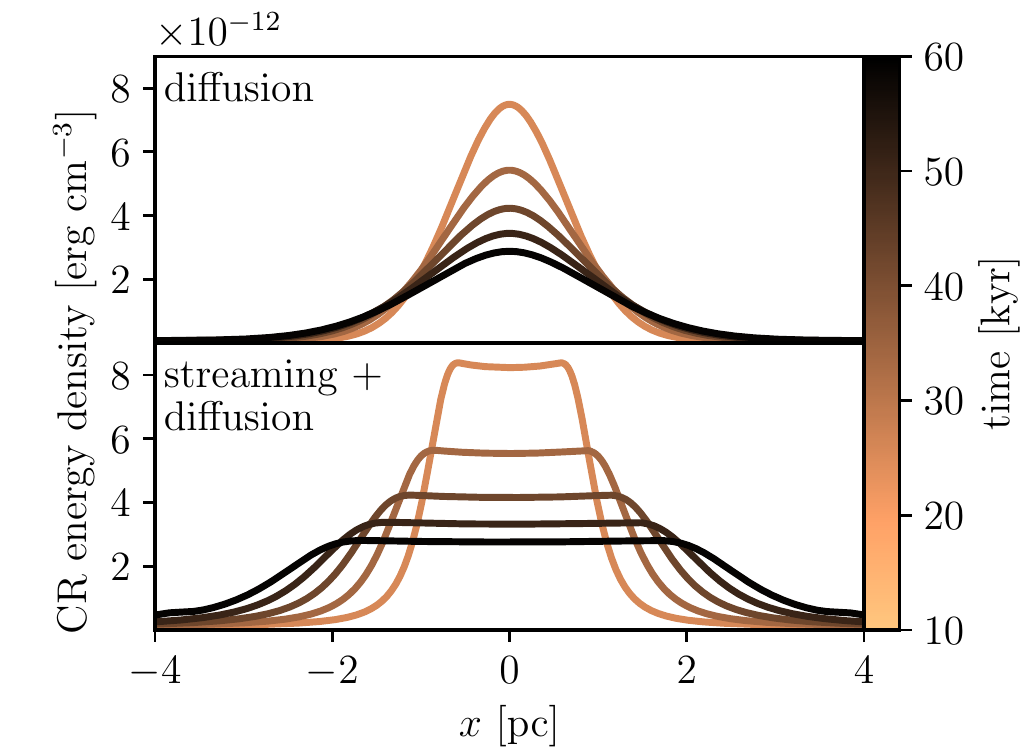}
    \caption{Evolution of the energy densities of CR electrons for the two propagation models over the course of $60$ kyr.}
    \label{fig:simulated_profiles_colors}
\end{figure}

We present the evolution of the CR electron energy density in Fig.~\ref{fig:simulated_profiles_colors}. The result for the \textit{diffusion} model resembles the typical evolution of a diffusion process: the initial Gaussian approximately maintains its shape while increasing its physical extent. The deviations from a pure diffusion profile are caused by CR energy losses due to Alfv\'en-wave cooling. 

Including the interactions between CRs and Alfv\'en waves allows CRs to enter the streaming mode of CR transport. Therein the two wings of the Gaussian are traveling at speeds of $\sim\pm{v}_\mathrm{a}$ in opposite directions. In between the two wings the CRs are rarefied causing the development of a plateau of almost constant energy density. At later times, the CRs are unable to maintain a high level of energy contained in Alfv\'en waves. As a result, CRs get less frequently scattered and enter the diffusive regime of CR transport.

\begin{figure*}
    \centering
    \includegraphics[width=\textwidth]{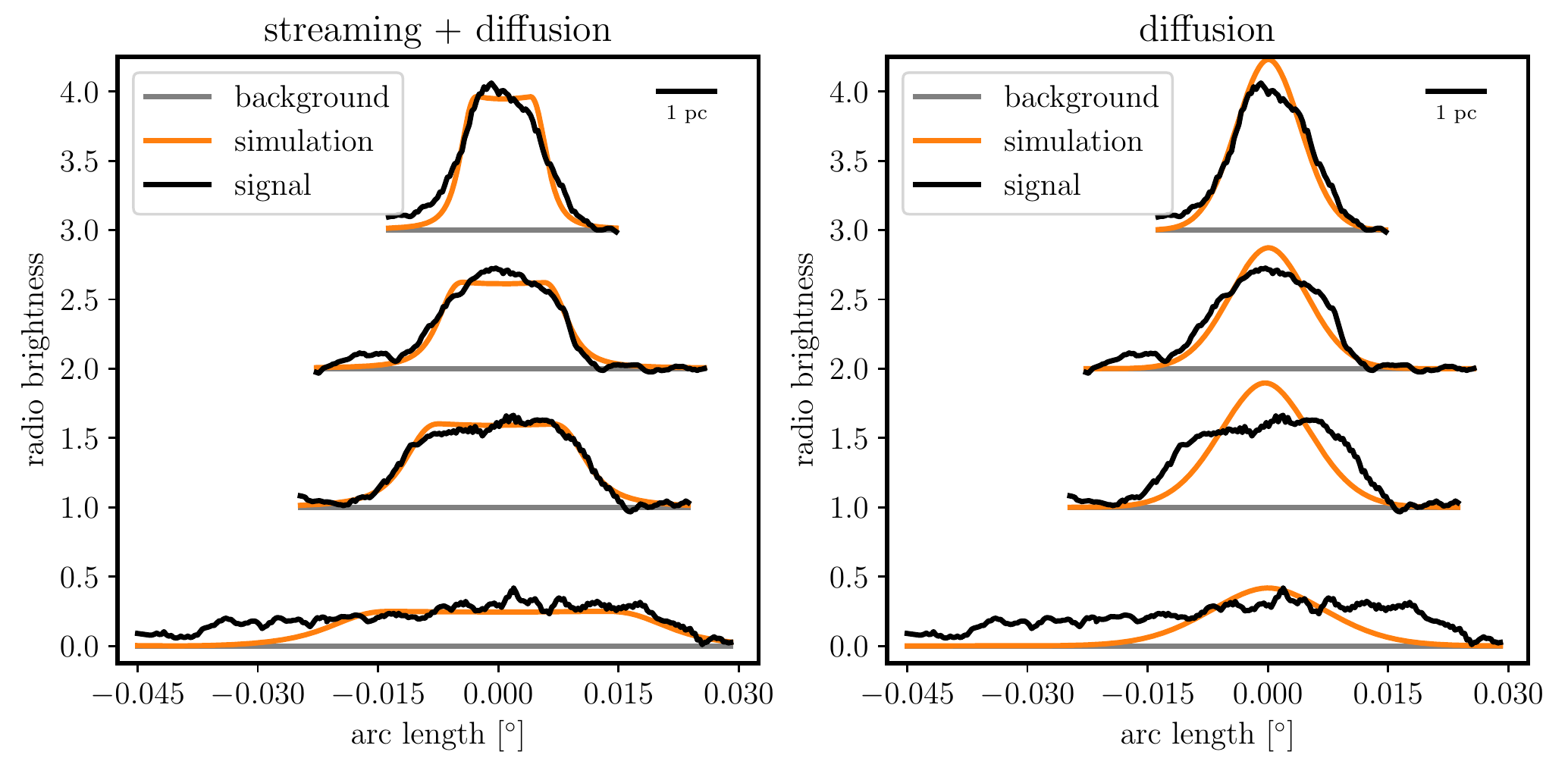}
    \caption{Comparison between the observed radio emission from the radio harp G359.47+0.12 \citep[extracted from][]{2019Heywood} and the simulated profiles. The simulated profiles are displayed at times $16.4$, $25.6$, $36.9$ and $71.8$~kyr (top to bottom) after CR injection. The filaments each have an offset of $1$ in the $y$-direction. We convert physical distances in the simulation to angular sizes assuming a distance of $8.2~\mathrm{kpc}$ to the CMZ. The \textit{streaming~+~diffusion} model matches the MeerKAT radio data significantly better than the \textit{diffusion} model.}
    \label{fig:lateral_cut_comp_full}
\end{figure*}

\section{Comparison to Observations}

We extract radio brightness profiles of the MeerKAT filaments \citep{2019Heywood} by taking cuts along the individual filaments of the harp G359.47+0.12, shown in Fig.~\ref{fig:harps} on the right. We use one segment for the three brightest filaments, respectively, and four segments for the faintest filament to trace its curvature. In Fig.~\ref{fig:lateral_cut_comp_full}, we compare this to our simulations by scaling the simulated CR energy density with a constant factor to match the observed radio flux. This factor is chosen so that the brightness in the first filament approximately agrees with the scaled simulated profiles. To match the brightness of the third and fourth filament in the \textit{streaming~+~diffusion} model, we had to increase the scaling by 25\% and 40\%, respectively. For the \textit{diffusion} model, we need to increase these factors by 50\%. The displayed background noise level is calculated by averaging the diffusive background excluding resolved and bright sources. 

Only the \textit{streaming~+~diffusion} model agrees with the observed profiles while the \textit{diffusion} model is unable the reproduce the central flat emission at late times because the diffusion profile never loses its central maximum. This causes a persistent overshoot in the very center and an underestimation of emission away so that the transition into the background is poorly modeled. The simulated Gaussian CR distribution progressively loses any steep gradients due to diffusion whereas the observations show continuously sharp transitions. Contrarily, both features can be nicely explained by including the effects of CR streaming as done in the \textit{streaming~+~diffusion} model. Therein, the flat emission naturally corresponds to the plateau of the rarefying CR energy density while the expanding fronts of the CR distribution follow the steep transition of the radio emission.   

There are no primary beam corrections applied to the four pointings that make up the MeerKAT mosaic \citep{2019Heywood}. While this precludes accurate photometry of the large-scale emission, the small-scale radio-harp profiles should mostly be unaffected. We note that the image of G359.47+0.12 shows a circularly shaped area with reduced flux levels of filaments and background emission, which is centered just outside the image in the lower right part of Fig.~\ref{fig:harps}. This reduced flux might be an artefact of the lacking primary beam corrections during imaging \citep{2019Heywood} and could explain the asymmetric shape of the older synchrotron filaments in Fig.~\ref{fig:lateral_cut_comp_full}. If correct, the agreement of the \textit{streaming~+~diffusion} model with the observation may improve even more after primary beam corrections and the \textit{diffusion} model will become worse, strengthening our finding.

We attempted to apply the same analysis to the harp G359.85+0.39. However, its filaments appear to be overlapping in projection. Whether the overlap is caused by the projection of individual spatially separated or of braided flux tubes that divert away from the central bright emission is not obvious. This ambiguity precludes a simple emission modeling of the complex flux-tube structure. However, the morphological similarity of both harps, which exhibit the shape of an inverted bell curve, strongly suggests that CR streaming is responsible for the emission structure in both cases. 

We predict a massive star or pulsar at the tip of each radio harp and encourage observers to search for them.

\section{Conclusions}
Here, we presented a model that explains the morphological appearance of the new phenomenon of radio harps observed within the bipolar outflows by MeerKAT. A careful modeling of two competing CR transport schemes (pure CR diffusion and a combination of CR streaming and diffusion in the self-confinement picture) demonstrates that only the CR \textit{streaming} model is able to match the detailed brightness distributions of the individual NTFs of the harp G359.47+0.12. The intermittency of the harp emission either reveals details of the magnetic field structure or about the magnetic reconnection processes at the interface of the shocked stellar (or pulsar) wind with the surrounding interstellar magnetic field. We hope that future high-resolution observations shed light on this, further enable us to disentangle the possibly three-dimensional structure of the other harp G359.85+0.39, and detect even more examples of this phenomenon. This will consolidate our conclusions regarding the CR streaming to be the relevant propagation mode for CRs at GeV energies.

\acknowledgments
TT and CP acknowledge support by the European Research Council under ERC-CoG grant CRAGSMAN-646955. This research was supported in part by the National Science Foundation under Grant No.\ NSF PHY-1748958.

\bibliography{main}

\end{document}